\documentclass[pra,twocolumn,tightenlines,showpacs,nofootinbib]{revtex4}
\usepackage{bm,dcolumn,amsmath,graphicx}

\newcommand{\eref}[1]{Eq.~(\ref{#1})}
\newcommand{\tref}[1]{Table~\ref{#1}}

\begin{document}

\title{Electronic bridge process in $^{229}$Th$^+$}

\author{S. G. Porsev$^{1,2}$}
\author{V. V. Flambaum$^1$}
\affiliation{$^1$ School of Physics, University of New South Wales,
Sydney, New South Wales 2052, Australia}
\affiliation{$^2$ Petersburg Nuclear Physics Institute, Gatchina,
Leningrad district, 188300, Russia}

\date{ \today }
\pacs{31.15.A-, 23.20.Lv, 27.90.+b}

\begin{abstract}
We have studied the  effect of atomic electrons on the nuclear transition
from the isomeric $^{229m}$Th state to the ground $^{229g}$Th state
in $^{229}$Th$^+$  due to the electronic bridge process.
The exact value of the nuclear transition frequency
is unknown so far; therefore, we have developed a formalism that can be
used for any nuclear transition frequency. We have calculated positions of
several high-lying even-parity states which are  not presented in experimental
 atomic spectra databases. We have found their energy levels and $g$ factors.
\end{abstract}

\maketitle

\section{Introduction}
\label{sec_I}
The energy splitting of the ground-state doublet of the $^{229}$Th nucleus
is only several electron volts~\cite{KroRei76}. At the same time the exact value
of the frequency for the transition from the isomeric $^{229m}$Th state
to the ground $^{229g}$Th state is unknown. Experiments give
 values of this frequency $\omega_N$ ranging from $3.5 \pm 1.0$ eV~\cite{ReiHel90}
to $7.6 \pm 0.5$ eV~\cite{BecBecBei07}. The measurements of
the lifetime of the isomeric state performed by different experimental groups
lead to values which differ from each other by many orders of magnitude
(see, e.g.,~\cite{InaHab09,MitHarOht03}).

As was noted in~\cite{PeiTam03} the nuclear transition from the isomeric state to the
ground state is of a great interest since it makes it possible to build a
very precise nuclear clock. This transition is very sensitive to hypothetical
temporal variation of the fundamental constants~\cite{Fla06}.

The triply ionized $^{232}$Th was recently laser cooled~\cite{CamSteChu09}.
Further, this experimental group plans to investigate the nuclear transition between
the isomeric and the ground state in a cold $^{229}$Th$^{3+}$ ion. Another experimental
group~\cite{Pei10} plans to use the ion $^{229}$Th$^+$ to study the nuclear
$^{229m}$Th--$^{229g}$Th transition.

In our previous work~\cite{PorFla10} we considered the $^{229}$Th$^{3+}$ ion and
calculated the transition probability of the $^{229}$Th nucleus
from its lowest energy isomeric state $^{229m}$Th to the ground state  $^{229g}$Th due to
the electronic bridge (EB) process. In this paper we consider the more complicated
three-valence ion $^{229}$Th$^+$. In our approach we do not fix the
value of the nuclear transition frequency. Hence, the result obtained here
can be applied for any value of $\omega_N$.

The paper is organized as follows.
In Sec.~\ref{sec_GF} we briefly discuss the general formalism describing
the EB process. In Sec.~\ref{sec_MC} we describe the method of calculation
of the properties of Th$^+$. Section~\ref{sec_RD} is devoted to the results
of calculations and Sec.~\ref{sec_C} contains concluding remarks.
Atomic units ($\hbar = |e| = m_e = 1$ and the speed of light $c = 137$)
are used throughout.
\section{General formalism}
\label{sec_GF}
A derivation of the equation for the probability of the EB process, $\Gamma_{\rm EB}$,
is given in detail in~\cite{PorFla10}. For this reason we will repeat here
only the main features of the formalism.

\begin{figure}
\includegraphics[scale=0.5]{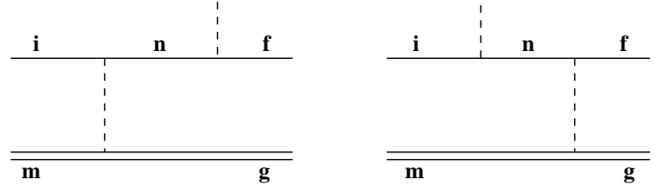}
\caption{Feynman diagrams of the EB process. The single and double solid lines relate to
the electronic and the nuclear transitions, correspondingly.
The dashed line is the photon line.}
\label{Fig:EB}
\end{figure}

The EB process can be represented by the two Feynman diagrams in Fig.~\ref{Fig:EB}.
In the following we assume that the initial $i$ and the final $f$ electronic states are of
opposite parity and fixed. A real photon which is emitted or absorbed is the electric dipole photon.
The EB process can be effectively treated as the electric
dipole $i \rightarrow f$ transition of the electron accompanied by the nuclear
transition from its isomeric state to the ground state.

Because the exact value of the nuclear transition frequency is unknown
 we do not fix it in our calculation.
Using the experimental data we suggest that most probably the real value
of $\omega_N$ is between 2 and 8 eV. The general expression for
$\Gamma_{\rm EB}$ we used for calculation of the EB process for
$^{229}$Th$^{3+}$ can be simplified for $^{229}$Th$^+$.
This is due to the spectrum of Th$^+$ being much denser than the spectrum
of Th$^{3+}$. As a result, for any nuclear transition frequency $\omega_N$
lying between 2 and 8 eV we can find an atomic transition from the initial
state $i$ to the definite intermediate state $n$ whose frequency will be very close
to $\omega_N$. Assuming the resonance character of the EB
process we arrive at the following expression for $\Gamma_{\rm EB}$~\cite{PorFla10}:
\begin{eqnarray}
\Gamma_{\rm EB} \approx \frac{4}{9}\left( \frac{\omega}{c}\right)^3
\frac{|\langle I_g||\mathcal{M}_1||I_m \rangle|^2}{(2I_m+1)(2J_i+1)}\, G_1 ,
\label{GamK}
\end{eqnarray}
where $\mathcal{M}_1$ is the magnetic dipole nuclear moment and
$|I_g \rangle $ and $|I_m \rangle$ are the ground nuclear state and
the isomeric nuclear state ($I_g = 5/2^+$, [633] Nilsson state and
$I_m = 3/2^+$, [631] Nilsson state);
$J_i$ is the electron total angular momentum of the initial state,
$\omega$ is the real photon frequency determined from the law of conservation of energy
as $\omega = \varepsilon_i - \varepsilon_f + \omega_N$ (where $\varepsilon_k$
is the atomic energy), and $G_1$ can be approximated by
\begin{eqnarray}
 G_1 \approx  \frac{1}{2J_n+1}  
 \left\vert \frac
{\langle \gamma_f J_f||D||\gamma_n J_n \rangle
 \langle \gamma_n J_n||\mathcal{T}_1|| \gamma_i J_i \rangle}
{\omega_{in}+\omega_N} \right\vert ^2.
\label{G1}
\end{eqnarray}
Here $\mathcal{T}_1$ is the electronic magnetic-dipole hyperfine coupling operator.
The total hyperfine coupling Hamiltonian $H_{\rm HFI}$ may be represented as
\begin{equation}
H_{\rm HFI} = \sum_\lambda \mathcal{M}_1^\lambda \, \mathcal{T}_{1 \lambda} .
\end{equation}
The operator $D$ is the electric dipole moment operator,
$\omega_{in} \equiv \varepsilon_i - \varepsilon_n$,
and $\gamma_k$ encapsulates all other electronic quantum numbers.
The explicit expressions for the matrix elements of the operators $\mathcal{T}_1$
and $D$ are given in our paper~\cite{PorFla10}.
If we introduce the quantity
\begin{eqnarray}
R_n \equiv |\langle \gamma_f J_f||D||\gamma_n J_n \rangle
         \langle \gamma_n J_n||\mathcal{T}_1|| \gamma_i J_i \rangle|^2
\label{Rn}
\end{eqnarray}
then \eref{G1} can be rewritten as
\begin{eqnarray}
 G_1 \approx  \frac{1}{2J_n+1}
 \frac{R_n}{(\omega_{in}+\omega_N)^2}.
\label{G1new}
\end{eqnarray}
In Eqs.~(\ref{G1})--(\ref{G1new}) the electronic state $|\gamma_n J_n \rangle$ is assumed
to be fixed. This state should be chosen to meet two conditions:
1) $-\omega_{in} \approx \omega_N$ and 2) if the first condition is fulfilled
for two atomic states we should take the state for which $R_n$ is larger.
The second condition is important because in certain cases the coefficients
$R_n$ for two neighboring energy levels differ by several orders of magnitude.

In Ref.~\cite{PorFla10} we used the dimensionless quantity
$\beta_{M1}$ defined as the ratio of the probability of the EB process,
$\Gamma_{\rm EB}$, to the probability of the $M1$ radiative nuclear
$m \rightarrow g$ transition, $\Gamma_N$:
\begin{eqnarray}
\beta_{M1} = \frac{\Gamma_{\rm EB}}{\Gamma_N}
\approx \left( \frac{\omega}{\omega_N} \right)^3 \frac{G_1}{3 (2J_i+1)} .
\label{betaM1}
\end{eqnarray}

It is reasonable to choose the ground state ($6d^2\, 7s$) $J=3/2$
as the initial state $i$ and consider the lowest lying odd-parity
state ($5f\, 7s^2$) $J=5/2$ as the final state $f$.
Thus, the intermediate atomic states contributing to $G_1$ are even-parity states
and our purpose is to calculate the coefficients $R_n$ for all
even-parity states whose transition frequencies to the ground state
are between 2 and 8 eV. Then, using Eqs.~(\ref{G1new}) and
(\ref{betaM1}) we can find the quantities $G_1$ and $\beta_{M1}$,
correspondingly, for any $\omega_N$ lying between 2 and 8 eV.
\section{Method of calculation}
\label{sec_MC}
We consider Th$^+$ as the atom with three valence
electrons above closed-shell core [1$s^2$, \ldots ,6$p^6$].
We employ the approach combining the
configuration-interaction (CI) method in the valence space
with many-body perturbation theory (MBPT) for core polarization
effects. In the following we refer to this combined approach as the
CI+MBPT method~\cite{DzuFlaKoz96b}.

At the first stage we have solved Dirac-Hartree-Fock (DHF)
equations~\cite{BraDeyTup77} in the $V^{N-3}$ approximation. This means
that the DHF equations were solved self-consistently for the core
electrons. After that we determined the $5f$, $6d$, $7p$, $7s$, and
$8s$ orbitals from the frozen-core DHF equations. The virtual
orbitals were determined with the help of a recurrent
procedure~\cite{KozPorFla96}. The one-electron basis set included
1$s$--18$s$, 2$p$--17$p$, 3$d$--16$d$, and 4$f$--15$f$ orbitals on
the CI stage.

The configuration spaces for even-parity and odd-parity states were
formed as follows. The main configuration of the ground state is
$6d^2 7s$. We formed the configuration space for the even-parity states
by allowing single, double, and triple excitations from the $6d^2 7s$
configuration to the 7$s$--13$s$, 7$p$--12$p$, 6$d$--11$d$, and 5$f$--10$f$
shells. The main configuration of the lowest lying odd-parity state is
$5f 7s^2$. The configuration space for the odd-parity levels was formed
by single, double, and triple excitations from the $5f 7s^2$
configuration to the 7$s$--13$s$, 7$p$--12$p$, 6$d$--11$d$, and 5$f$--10$f$
shells. Inclusion of all possible (up to triple) excitations is important,
especially for high-lying states. It allows us to take into
account most completely the configuration interaction for all
considered states.

In the CI+MBPT method, the energies and the wave functions are determined
from the eigenvalue equation in the model space of the valence electrons,
\begin{equation}
H_{\mathrm{eff}}(E_{p})\,|\Phi _{p}\rangle =E_{p}\,|\Phi _{p}\rangle \,,
\label{Eqn_Sh}
\end{equation}
where the effective Hamiltonian is defined as
\begin{equation}
H_{\mathrm{eff}}(E)=H_{\mathrm{FC}}+\Sigma (E).  \label{Eqn_Heff}
\end{equation}
Here $H_{\mathrm{FC}}$ is the relativistic three-electron Hamiltonian in the
frozen-core approximation and $\Sigma(E)$ is the energy-dependent
core-polarization correction.

Together with the effective Hamiltonian $H_{\rm eff}$ we
introduce the effective electric-dipole operator $D_{\rm eff}$ and
the operator $(\mathcal{T}_1)_{\rm eff}$ acting in the model
space of valence electrons. These operators were obtained within
the  relativistic random-phase approximation
(RPA)~\cite{DzuKozPor98,KolJohSho82}, which
describes a shielding of the externally applied
electric field by the core electrons. The RPA sequence
of diagrams was summed to all orders of the perturbation theory.

To solve the RPA equations and to calculate diagrams for the effective Hamiltonian and
the effective operators $D$ and $\mathcal{T}_1$ we used a different basis set. The core
orbitals in this basis set are the same as before, but the
number of virtual orbitals is much larger.  On the whole, it consisted
of $1s$--$22s$, $2p$--$22p$, $3d$--$22d$, $4f$--$22f$, and
$5g$--$16g$ orbitals.

\section{Results and discussion}
\label{sec_RD}
We start the discussion of the results with the following remark:
The spectrum of Th$^+$ is very complicated. As is seen
from the experimental data~\cite{ThII}, on the one hand, the
states belonging to different configurations strongly interact with each
other and $LS$ coupling is not valid (even approximately) for this ion.
On the other hand, it is not a chaotic system. Respectively,
the methods of statistical physics are not applicable.
Such an ``intermediate'' type of coupling makes the calculations of the
properties of Th$^+$ rather difficult.

As we have already mentioned in Sec.~\ref{sec_GF}, we consider the following
transition:
$6d^2 7s\, (J=3/2) \stackrel{\mathcal{T}_1}{\longrightarrow} n
                   \stackrel{E1}{\longrightarrow} 5f 7s^2\, (J=5/2)$.
According to Eqs.~(\ref{G1}) -- (\ref{G1new}) only intermediate states $n$
with $J_n=3/2$ and $J_n=5/2$ contribute to the probability of the EB
process for this transitions.

In Tables~\ref{Tab:1} and \ref{Tab:2} we presented the calculated
values of the energy levels with $J_n=3/2$ and $J_n=5/2$ and also
$g$ factors and the coefficients $R_n$
obtained with use of \eref{Rn} for the most interesting frequency range from
2 to 8 eV. In Table~\ref{Tab:1} we present the results for the atomic
frequencies from 2 to 5 eV and in Table~\ref{Tab:2} (which is a continuation
of Table~\ref{Tab:1}) the data are listed for the frequencies from 5 to 8 eV.
The results for the energy levels and $g$ factors were obtained in the CI+MBPT approximation.
The values of the coefficients $R_n$ were found in the frame of the CI+MBPT+RPA approach.
\begin{table}
\caption{The low-lying energy levels in the range from 18119 to 40644 cm$^{-1}$
(from 2 to 5 eV) in the CI+MBPT approximation, $g$ factors, and the coefficients
$R_n$ (in a.u.). $\Delta$ is the difference between the energies of the
ground state and the excited state. The notation $y[x]$ means $y \times 10^x$.
The theoretical values are compared with the experimental data.}

\label{Tab:1}

\begin{ruledtabular}
\begin{tabular}{lcccccc}
\multicolumn{1}{c}{Conf.} & \multicolumn{1}{c}{$J$}
& \multicolumn{1}{c}{$\Delta$(exp)\footnotemark[1]} & \multicolumn{1}{c}{$\Delta$(calc)}
& \multicolumn{1}{c}{g(exp)\footnotemark[1]} & \multicolumn{1}{c}{g(calc)}
& \multicolumn{1}{c}{$R_n$} \\
$6d^2\,7s $ & 3/2 &   0   &   0   & 0.639 &  0.712  &  9[-2] \\
$6d^3     $ & 3/2 & 18119 & 21351 & 0.93  &  0.887  &  4[-3] \\
$6d^3     $ & 5/2 & 20159 & 23731 & 1.19  &  1.198  &  2[-3] \\
$6d^3     $ & 5/2 & 22106 & 26005 & 0.92  &  0.931  &  3[-3] \\
$6d^3     $ & 3/2 & 25382 & 29632 & 1.25  &  1.242  &  2[-7] \\
$5f7s7p   $ & 5/2 & 26489 & 26971 & 0.776 &  0.747  &  5[-3] \\
$5f^2\,7s $ & 3/2 & 26762 & 27561 & 0.4   &  0.480  &  1[-3] \\
$5f^2\,7s $ & 5/2 & 27594 & 28396 & 0.963 &  0.975  &  1[-6] \\
$5f7s7p   $ & 3/2 & 27631 & 28082 & 0.625 &  0.518  &  3[-2] \\
$6d^3     $ & 3/2 & 28011 & 32348 & 0.717 &  0.841  &  1[-6] \\
$6d^3     $ & 5/2 & 28026 & 32764 & 1.13  &  0.975  &  1[-4] \\
$5f7s7p   $ & 5/2 & 28824 & 29367 & 0.987 &  0.993  &  2[-1] \\
$5f^2\,7s $ & 5/2 & 29346 & 30440 & 0.935 &  0.933  &  1[-5] \\
$5f6d7p   $ & 5/2 & 31259 & 31973 & 0.781 &  0.903  &  2[-3] \\
$5f7s7p   $ & 5/2 & 31754 & 32554 & 0.948 &  0.997  &  4[-3] \\
$5f6d7p   $ & 3/2 & 32959 & 34051 & 0.874 &  0.834  &  2[-4] \\
$5f^2 7s  $ & 5/2 & 33731 & 34891 & 1.031 &  1.014  &  3[-2] \\
$5f^2 7s  $ & 3/2 & 34019 & 35306 & 0.823 &  0.910  &  4[-2] \\
$5f^2 7s  $ & 5/2 & 34175 & 35727 & 0.986 &  1.095  &  1[-2] \\
$5f7s7p   $ & 5/2 & 34544 & 34732 & 1.003 &  0.965  &  2[-2] \\
$5f7s7p   $ & 3/2 & 35021 & 35535 & 1.042 &  1.001  &  2[-3] \\
$5f6d7p   $ & 5/2 & 35741 & 37326 & 0.954 &  0.996  &  8[-3] \\
$5f^2 6d  $ & 5/2 & 36066 & 36864 & 0.887 &  0.834  &  6[-2] \\
$5f^2 7s  $ & 3/2 & 36329 & 38069 & 1.615 &  1.636  &  4[-6] \\
$5f6d7p   $ & 5/2 & 37465 & 39615 & 1.048 &  0.958  &  3[-3] \\
$5f^2 7s  $ & 3/2 & 37542 & 38787 & 1.003 &  0.870  &  1[-2] \\
$5f7s7p   $ & 3/2 & 37822 & 39376 & 1.15  &  0.942  &  1[-2] \\
$5f7s7p   $ & 5/2 & 37945 & 38975 & 0.893 &  0.987  &  3[-2] \\
$5f^2 7s  $ & 5/2 & 38105 & 38653 & 1.172 &  0.918  &  3[-2] \\
$5f^2 6d  $ & 3/2 & 38372 & 40194 & 1.200 &  1.295  &  1[-3] \\
$5f6d7p   $ & 5/2 & 38729 & 40305 & 1.255 &  1.177  &  3[-4] \\
$5f^2 7s  $ & 3/2 & 38757 & 40937 & 0.935 &  0.903  &  2[-4] \\
$5f6d7p   $ & 3/2 & 38836 & 39919 & 1.013 &  1.046  &  2[-4] \\
$5f6d7p   $ & 5/2 & 38864 & 40386 & 0.967 &  1.076  &  4[-6] \\
$5f6d7p   $ & 3/2 & 39151 & 40336 & 0.739 &  0.823  &  4[-3] \\
$5f7s7p   $ & 5/2 & 39367 & 40679 & 1.140 &  1.277  &  7[-3] \\
$5f6d7p   $ & 5/2 & 39701 & 41000 & 1.090 &  1.177  &  2[-3] \\
$5f6d7p   $ & 5/2 & 40216 & 41967 & 1.024 &  0.941  &  2[-3] \\
$5f7s7p   $ & 3/2 & 40223 & 41738 & 0.738 &  0.887  &  7[-4] \\
$5f7s7p   $ & 3/2 & 40278 & 41890 & 0.705 &  0.595  &  7[-4] \\
$5f^2 6d  $ & 5/2 & 40644 & 42229 & 0.856 &  0.979  &  1[-3]
\end{tabular}
\end{ruledtabular}
\footnotemark[1]{Reference~\cite{ThII}}. \\
\end{table}
\begingroup
\squeezetable
\begin{table*}
\caption{The low-lying energy levels in the range from 40924 to 64000 cm$^{-1}$
(from 5 to 8 eV) in the CI+MBPT approximation, $g$ factors, and the coefficients
$R_n$ (in a.u.). $\Delta$ is the difference between the energies of the
ground state and the excited state. The notation $y[x]$ means $y \times 10^x$.}

\label{Tab:2}

\begin{ruledtabular}
\begin{tabular}{lcccccc}
\multicolumn{1}{c}{Conf.} & \multicolumn{1}{c}{$J$}
& \multicolumn{1}{c}{$\Delta$(exp)\footnotemark[1]} & \multicolumn{1}{c}{$\Delta$(calc)}
& \multicolumn{1}{c}{g(exp)\footnotemark[1]} & \multicolumn{1}{c}{g(calc)}
& \multicolumn{1}{c}{$R_n$} \\
\hline
$5f6d7p        $ & 5/2 & 40924 & 42449 & 0.988 &  1.063  &  3[-3] \\
$5f6d7p        $ & 3/2 & 40992 & 42501 & 1.036 &  1.037  &  3[-5] \\
$5f7s7p        $ & 5/2 & 41328 & 42747 & 1.101 &  0.957  &  1[-4] \\
$5f^2 6d       $ & 3/2 & 41677 & 43943 & 1.220 &  1.264  &  1[-6] \\
$5f^2 6d       $ & 3/2 & 41937 & 43236 & 1.095 &  1.088  &  1[-4] \\
$5f^2 6d+5f6d7p$ & 5/2 & 42337 & 44239 & 1.15  &  1.041  &  7[-4] \\
$5f^2 6d+5f6d7p$ & 5/2 & 42352 & 44305 & 1.126 &  1.237  &  5[-4] \\
$5f^2 6d+5f6d7p$ & 5/2 & 43097 & 44716 & 0.982 &  0.995  &  2[-3] \\
$5f^2 6d+5f6d7p$ & 5/2 & 43228 & 44876 & 1.153 &  1.135  &  1[-6] \\
$5f^2 6d       $ & 3/2 & 43245 & 45161 & 1.08  &  1.107  &  1[-5] \\
$5f^2 6d+5f6d7p$ & 5/2 & 43772 & 45900 & 1.04  &  0.985  &  3[-3] \\
$5f6d7p        $ & 3/2 & 43808 & 45544 & 1.211 &  1.271  &  1[-5] \\
$5f^2 6d       $ & 3/2 & 44301 & 46287 & 1.342 &  1.357  &  1[-5] \\
$5f^2 6d+5f6d7p$ & 5/2 & 44389 & 46283 & 1.158 &  1.087  &  1[-3] \\
$5f^2 6d+5f6d7p$ & 5/2 & 44553 & 46775 & 1.182 &  1.224  &  2[-6] \\
$5f^2 6d       $ & 3/2 & 44890 & 46742 & 1.346 &  0.960  &  3[-5] \\
$5f^2 6d+5f6d7p$ & 5/2 & 45190 & 46928 & 0.674 &  0.729  &  4[-5] \\
$5f6d7p        $ & 3/2 & 45306 & 46994 & 0.6   &  0.910  &  8[-4] \\
$5f^2 6d+5f6d7p$ & 5/2 & 45611 & 47310 & 1.075 &  1.076  &  1[-8] \\
$5f^2 6d+5f6d7p$ & 5/2 & 45800 & 47877 & 1.3   &  1.249  &  5[-4] \\
$5f6d7p        $ & 3/2 & 46264 & 47778 & 0.891 &  0.936  &  2[-3] \\
$5f6d7p        $ & 3/2 & 46396 & 48554 &       &  1.268  &  1[-4] \\
$5f^2 6d+5f6d7p$ & 5/2 & 46581 & 48439 & 1.018 &  1.058  &  2[-3] \\
$5f^2 6d+5f6d7p$ & 5/2 & 46603 & 48616 & 1.112 &  1.135  &  2[-4] \\
$5f^2 6d+5f6d7p$ & 5/2 & 46903 & 48835 & 1.143 &  1.147  &  1[-3] \\
$5f^2 6d       $ & 3/2 & 46936 & 49401 & 0.956 &  0.567  &  6[-4] \\
$5f6d7p        $ & 3/2 & 47149 & 49137 & 1.09  &  1.316  &  1[-4] \\
$5f^2 6d+5f6d7p$ & 5/2 & 47324 & 49355 & 1.189 &  1.231  &  1[-4] \\
$5f^2 6d       $ & 3/2 & 47870 & 50324 &       &  0.849  &  4[-4] \\
$5f^2 6d+5f6d7p$ & 5/2 & 48321 & 50553 &       &  1.155  &  2[-4] \\
$5f^2 6d+5f6d7p$ & 5/2 & 48492 & 50633 &       &  1.025  &  2[-4] \\
$5f6d7p        $ & 3/2 & 48690 & 50924 & 0.922 &  1.079  &  2[-5] \\
$5f^2 6d       $ & 3/2 & 48818 & 50749 & 0.956 &  0.727  &  2[-4] \\
$5f^2 6d+5f6d7p$ & 5/2 & 49069 & 51463 &       &  1.061  &  4[-7] \\
$5f6d7p        $ & 3/2 & 49415 & 51692 & 1.003 &  1.213  &  1[-6] \\
$5f^2 6d+5f6d7p$ & 5/2 & 49873 & 51941 &       &  1.054  &  5[-4] \\
$5f^2 6d+5f6d7p$ & 5/2 & 50664 & 52964 &       &  1.207  &  2[-4] \\
$5f 6d 7p      $ & 3/2 & 50735 & 52761 & 1.36  &  1.585  &  1[-7] \\
$5f 6d 7p      $ & 3/2 & 50908 & 53760 & 1.3   &  0.852  &  3[-4] \\
$5f^2 6d+5f6d7p$ & 3/2 & 51025 & 54511 & 1.270 &  1.286  &  2[-5] \\
$5f^2 6d+5f6d7p$ & 5/2 & 51363 & 54363 &       &  1.271  &  3[-4] \\
$5f6d7p        $ & 3/2 & 51676 & 54796 &       &  1.069  &  3[-4] \\
$5f^2 6d+5f6d7p$ & 5/2 & 51865 & 54851 &       &  1.031  &  1[-3] \\
$5f^2 6d+5f6d7p$ & 5/2 & 51936 & 55511 &       &  1.279  &  1[-3] \\
$5f6d7p        $ & 3/2 & 52307 & 55562 &       &  1.036  &  2[-3] \\
$5f6d7p        $ & 3/2 & 52736 & 57665 &       &  1.246  &  3[-3] \\
$5f^2 6d+5f6d7p$ & 5/2 & 53845 & 56279 &       &  1.253  &  1[-4] \\
$5f^2 6d+5f6d7p$ & 5/2 & 54494 & 57274 &       &  1.198  &  2[-3] \\
$5f6d7p        $ & 3/2 & 54922 & 58868 &       &  1.102  &  6[-6] \\
$5f6d7p        $ & 3/2 & 56235 & 59107 &       &  0.884  &  2[-5] \\
$5f^2 6d+5f6d7p$ & 5/2 & 56391 & 58037 &       &  1.446  &  1[-4] \\
$6d^2 8s       $ & 3/2 &       & 58119 &       &  0.645  &  6[-3] \\
$6d^2 8s       $ & 5/2 &       & 58301 &       &  1.073  &  2[-2] \\
$5f^2 6d+5f6d7p$ & 5/2 &       & 59731 &       &  0.932  &  1[-1] \\
$6d7s8s        $ & 3/2 &       & 59808 &       &  1.092  &  3[-3] \\
$5f6d7p        $ & 3/2 &       & 60287 &       &  1.167  &  1[-3] \\
$6d7s8s        $ & 5/2 &       & 60416 &       &  1.205  &  2[-2] \\
$5f6d7p        $ & 5/2 &       & 60462 &       &  1.216  &  1[-3] \\
$6d7s8s        $ & 3/2 &       & 61763 &       &  0.784  &  1[-6] \\
$5f^2 6d       $ & 5/2 &       & 61996 &       &  1.183  &  9[-5] \\
$6d^2 8s       $ & 5/2 &       & 62345 &       &  0.937  &  2[-5] \\
$6d7s8s        $ & 3/2 &       & 62927 &       &  0.839  &  9[-4] \\
$6d^2 7d       $ & 5/2 &       & 63308 &       &  0.857  &  2[-4] \\
$6d^2 7d       $ & 3/2 &       & 63381 &       &  1.079  &  1[-6] \\
$6d^2 7d       $ & 3/2 &       & 63729 &       &  0.928  &  3[-5] \\
$6d^2 8s+6d7s8s$ & 5/2 &       & 63955 &       &  1.065  &  7[-5] \\
\end{tabular}
\end{ruledtabular}
\footnotemark[1]{Reference~\cite{ThII}}. \\
\end{table*}
\endgroup

As is seen from the tables basically the agreement between the experimental
and the calculated energy levels is satisfactory. For the majority
of the levels presented in Tables~\ref{Tab:1} and \ref{Tab:2} the agreement
is at the level of several percent. The largest difference between
the experimental and the theoretical values is for the states belonging to the
$6d^3$ configuration, where it reaches 15\%. At the same time the $g$ factors for
these states were reproduced rather well. This means that the configuration
interaction was taken into account correctly.

The energy levels with total angular momenta $J=3/2$ and $J=5/2$ lying higher
than 56391 cm$^{-1}$ are not identified experimentally.
In our work we have determined several new high-lying energy levels with $J=3/2$
and 5/2. In the first rows of Tables~\ref{Tab:1} and \ref{Tab:2}
we indicate the configurations that give the largest contributions to these states
according to our calculation.

As we have already mentioned the configuration mixture is strong for all
states starting from the ground state. Sometimes we were unable to reproduce
correctly the configuration interaction. In such cases the theoretical $g$
factors differ from the experimental $g$ factors and, respectively, the
accuracy of calculation of $R_n$ for such states is poorer.

As follows from Tables~\ref{Tab:1} and \ref{Tab:2} the coefficients $R_n$
change from $10^{-7}$ to $10^{-1}$. This is not surprisingly if we note that
the initial state $6d^2 7s$ and the final state $5f 7s^2$ differ from each
other by two electrons while $T_1$ and $D$ are the one-electron operators.
For this reason the $ i \rightarrow n \rightarrow f$ transition occurs only by
the configuration interaction. In a case when the intermediate state $n$
is characterized by configurations that open two strong one-electron
$6d^2 7s \rightarrow n$ and $n \rightarrow 5f 7s^2$ transitions, $R_n$ turn
out to be large. Due to the complexity of the energy level spectrum of Th$^+$
the accuracy of the calculation of the coefficients $R_n$ is not high. We would
consider these values as an order-of-magnitude estimate.

To illustrate how the developed formalism works we consider two possible
values of the nuclear frequency, $\omega_N = 3.5 \, {\rm eV}$~\cite{ReiHel90}
and $\omega_N = 5.5 \, {\rm eV}$~\cite{GuiHel05},
as reported by two experimental groups in the mentioned papers.
In \tref{Tab:3} we present the values of the relevant quantities.

For $\omega_N = 3.5 \, {\rm eV} \approx 28231\, {\rm cm}^{-1}$
the resonance contribution to $\Gamma_{\rm EB}$ comes from the atomic state
$J=5/2$ at 28824 cm$^{-1}$ belonging to the configuration $5f 7s 7p$.
We chose this state because the transition frequency $\omega_{\rm res}$
from this state to the initial state $i$ (the ground state) is close to
$\omega_N$ and the coefficient $R_n$ is largest. Knowing from \tref{Tab:1} the
coefficient $R_n$ for this state and using Eqs.~(\ref{G1new}) and
(\ref{betaM1}) we can easily find the quantities $G_1$ and $\beta_{M1}$
for the transition $6d^2 7s\, (J=3/2) \stackrel{\mathcal{T}_1}{\longrightarrow} 5f 7s 7p\,(J=5/2)
\stackrel{E1}{\longrightarrow} 5f 7s^2\,(J=5/2)$.
In a similar way $G_1$ and $\beta_{M1}$ can be obtained for
$\omega_N = 5.5 \, {\rm eV}$.

Comparing the coefficients $\beta_{M1}$ obtained for $\omega_N = 3.5 \, {\rm eV}$
and $\omega_N = 5.5 \, {\rm eV}$ we see that they are of the order of $10^2$--$10^3$.
We note that in the case of $\omega_N = 5.5 \, {\rm eV}$
the difference $(\omega_{\rm res} - \omega_N)$ is only 26 cm$^{-1}$ while
$R_n = 0.001$ is rather small.
For $\omega_N = 3.5 \, {\rm eV}$ the difference
$(\omega_{\rm res} - \omega_N) \sim 600\, {\rm cm}^{-1}$
but the coefficient $R_n=0.2$ is two orders of magnitude
larger than that for $\omega_N = 3.5 \, {\rm eV}$.
The latter occurs because the resonance energy level whose frequency is close to
$\omega_N = 3.5 \, {\rm eV}$ belongs to the configuration $5f7s7p$. Hence, there is
a strong $5f7s7p\, (J=5/2) \stackrel{E1}{\longrightarrow} 5f 7s^2\, (J=5/2)$
 transition. Due to an admixture
of the configuration $6d^2 7s$ to the configuration $5f7s7p$ the amplitude of the
$6d^2 7s\, (J=3/2) \stackrel{\mathcal{T}_1}{\longrightarrow} 5f7s7p\, (J=5/2)$
transition is not small. As a result, the coefficient $R_n$ is large.

The case of $\omega_N = 7.6 \, {\rm eV}$~\cite{BecBecBei07} requires
special attention. The problem is that the atomic energy levels
are not identified experimentally in the region of 7.5 eV and,
consequently, we cannot compare the theoretical energy levels with the
experimental energy levels. As we previously mentioned, the theoretical accuracy
is at the level of several percent. Thus at present we are unable to reliably predict
the position of the resonance energy level and, consequently, the coefficient $\beta_{M1}$.
For this reason experimental investigations and identification of the energy
levels in the frequency region $\sim7.5 \, {\rm eV}$ would be very useful.
Once these tasks are completed the coefficient $\beta_{M1}$ can be easily determined.

\begin{table*}
\caption{The nuclear transition frequency $\omega_N$ (given in eV and in cm$^{-1}$)
along with the configuration, the total angular momentum $J$, and the transition frequency
with respect to the ground state ($\omega_{\rm res}$) for the resonance state mainly
contributing to $G_1$, listed with the coefficients $R_n$ (in a.u.),
$G_1$ (in a.u.), and $\beta_{M1}$.}

\label{Tab:3}

\begin{ruledtabular}
\begin{tabular}{lclcclrr}
\multicolumn{2}{c}{$\omega_N$} & \multicolumn{3}{c}{Resonance state} &
\multicolumn{1}{c}{}      & \multicolumn{1}{c}{}    & \multicolumn{1}{c}{} \\
\multicolumn{1}{c}{eV}    & \multicolumn{1}{c}{cm$^{-1}$} &
\multicolumn{1}{c}{Conf.} & \multicolumn{1}{c}{$J$} &
\multicolumn{1}{c}{$\omega_{\rm res}$(cm$^{-1}$)}   &\multicolumn{1}{c}{$R_n$} &
\multicolumn{1}{c}{$G_1$} & \multicolumn{1}{c}{$\beta_{M1}$} \\
\hline
3.5 & 28231 & $5f7s7p$           & 5/2 & 28824 & 0.2   &  4570  & 225 \\
5.5 & 44363 & $5f^2 6d + 5f6d7p$ & 5/2 & 44389 & 0.001 & 11880  & 720 \\
\end{tabular}
\end{ruledtabular}
\end{table*}
\section{Conclusion}
\label{sec_C}
To conclude, we have found several high-lying even-parity states
with total angular momenta $J=3/2$ and $J=5/2$ that are not identified
in the atomic spectra database~\cite{ThII}. We have
determined the energy levels and the $g$ factors of these states.

We have calculated the coefficients $R_n$ determined by \eref{Rn} for
the even-parity states lying between 2 and 8 eV. When the nuclear transition
frequency $\omega_N$ is exactly known and the atomic energy levels are
experimentally identified, we can find, using $R_n$, the
coefficients $G_1$, $\beta_{M1}$, and the probability of the EB process.
\section{Acknowledgments}
\label{sec_Ac}
We would like to thank E.~Peik for stimulating
discussion and interest to this work. This work was supported by
the Australian Research Council. The work of S.G.P. was supported in part by
the Russian Foundation for Basic Research under Grant No. 08-02-00460-a.


\end{document}